\documentclass[aps,prd,twocolumn,showpacs,superscriptaddress,nofootinbib]{revtex4-2}
\pdfoutput=1
\usepackage{graphicx}
\usepackage{bm}
\usepackage{amssymb,amsmath,latexsym}
\usepackage{color}
\definecolor{darkblue}{RGB}{0,0,196}
\definecolor{darkred}{RGB}{190,45,30}
\definecolor{violet}{RGB}{155,38,182}
\definecolor{magenta}{RGB}{255,0,255}
\usepackage[colorlinks=true,linktocpage=true,linkcolor=darkblue,citecolor=darkblue,urlcolor=darkblue]{hyperref}
\usepackage{array}
\usepackage{subfigure}

\begin{document}

\title{Axion effects in the stability of hybrid stars}

\author{Bruno S. Lopes} 
\affiliation{Departamento de F\'{\i}sica, Universidade Federal de
  Santa Maria, Santa Maria, RS 97105-900, Brazil}
  
\author{Ricardo L. S. Farias} \email{ricardo.farias@ufsm.br}
\affiliation{Departamento de F\'{\i}sica, Universidade Federal de
  Santa Maria, Santa Maria, RS 97105-900, Brazil}
  
\author{Veronica Dexheimer}
\affiliation{Department of Physics, Kent State University, Kent, OH 44242, USA}  
  
\author{Aritra Bandyopadhyay}
\affiliation{Guangdong Provincial Key Laboratory of Nuclear Science,
  Institute of Quantum Matter, South China Normal University,
  Guangzhou 510006, China}
\affiliation{Institut für Theoretische Physik, Universität Heidelberg, Philosophenweg 16, 69120 Heidelberg, Germany}

\author{Rudnei O. Ramos}
\affiliation{Departamento de F\'{\i}sica Te\'orica, Universidade do
  Estado do Rio de Janeiro, 20550-013 Rio de Janeiro, RJ, Brazil}


\begin{abstract}
We investigate the effects of including strong charge-parity (CP) violating effects through axion fields in the microscopic equation of state of massive hybrid neutron stars. We assume that their cores contain deconfined quark matter and include the effects of axions via an effective 't Hooft determinant interaction. The hadronic crusts are described using different approaches in
order to make our results more general. 
We find that the presence of axions stabilizes massive
hybrid neutron stars against gravitational collapse by weakening the deconfinement phase
transition and bringing it to lower densities. This enables to reproduce hybrid neutron stars
in agreement with modern astrophysical constraints.
\end{abstract}

\keywords{Quark-gluon plasma, Relativistic heavy-ion collisions,
  Quantum chromodynamics, Hard-thermal-loop, QCD phase diagram}

\maketitle


Recent developments in the field of observational
astronomy made possible through gravitational wave interferometers, along with the
Neutron Star Interior Composition Explorer
(NICER)~\cite{LIGOScientific:2018cki,Miller:2019cac,Miller:2021qha,Riley:2019yda,Riley:2021pdl}
have played a key role in providing tight constraints on neutron star (NS)
masses and radii. Consequently, the equation of state (EoS), which is
the most important ingredient in the characterization of strongly
interacting dense matter, has also been tightly constrained. Now that
the field of view is being narrowed down, we need powerful tools to
microscopically study the properties of massive NSs and, by doing
that, explore the dense region of the Quantum Chromodynamics (QCD)
phase diagram, which cannot be explored with current state of the
art terrestrial experiments and lattice QCD simulations.

The axion has long been considered as a prime constituent of cold
dark matter~\cite{Preskill:1982cy,Abbott:1982af,Dine:1982ah}. For a recent review, see e.g.,
Ref.~\cite{Chadha-Day:2021szb} and references therein. The QCD axion
and axion-like particles that are predicted to exist in extensions of
the standard model of particle physics are assumed to be an extremely light pseudo
Nambu-Goldstone boson, which couples very weakly to standard
hadronic matter~\cite{Peccei:1977hh,Marsh:2015xka}. The concept of
axions originated as the most appropriate solution to the problem of violation of combined symmetries of charge conjugation and parity (charge-parity, CP) in
QCD~\cite{Weinberg:1977ma,Wilczek:1977pj} and, since then, it has been
associated with various strongly interacting phenomena. The particular
motivation for studying the effects of axions on stellar objects
(including massive NSs) comes from the idea that they could take part in energy transport, and thus affect their thermal evolution~\cite{Raffelt:2006cw,Giannotti:2015kwo,Keller:2012yr,Sedrakian:2015krq,Sedrakian:2018kdm,Harris:2020qim}.
Axions, as prime dark matter candidates, may also influence neutron
star properties due to their possible continual accumulation
and by their gravitational capture during stellar formation. Hence,
NSs may contain a substantial amount of dark matter and, in
particular, dark matter in the form of axions. Dark matter in the form
of self-interacting bosonic particles has been studied recently in
connection to several properties of neutron
stars~\cite{Karkevandi:2021ygv}, and it was shown that this can affect their maximum mass and tidal
deformability. Axions, as bosons, are expected to share many of those
properties and, hence, lead to similar effects. Changes in neutron-star
composition affect the EoS,
thus influencing stellar stability, central density, and radius. This is
the subject we explore in the present paper, where we also consider different fermionic descriptions and interactions. 


The QCD axion has been recently studied in a hot and
magnetized medium in the context of the Nambu-Jona-Lasinio (NJL) model for quark matter~\cite{Bandyopadhyay:2019pml}. The NJL
model has been extensively used in the similar context of spontaneous
CP
violation~\cite{Fukushima:2001hr,Frank:2003ve,Boer:2008ct,Boomsma:2009eh,Boomsma:2009yk,Chatterjee:2014csa}. It
incorporates the effects of axions via an effective 't Hooft
determinant interaction between the quarks~\cite{tHooft:1976snw,tHooft:1986ooh}. In our
case, we are dealing with a much smaller energy scale than the axion
symmetry breaking energy (of the order of the scale in grand unified theories, $\sim 10^{15}$ GeV) and, hence, we can safely take the axion
field $a$ to be in its vacuum expectation value. Thus, the
Lagrangian density of the three quark flavor NJL model, including the
CP violating effects~\cite{Chatterjee:2011yz} through axion fields,
can be expressed in the following form for a quark of flavor $j=$ u,~d,~s:
\begin{align}
\mathcal{L} &= \bar{\psi_j}\left(i \gamma^{\mu} \partial_{\mu} -
m_{0}^j\right)\psi_j + G_s \sum_{b = 0}^{8}
\left[\left(\bar{\psi_j} \lambda^b \psi_j\right)^2 \right. \nonumber \\ &\left.+ \left(\bar{\psi_j}i
  \gamma_5 \lambda^b \psi_j\right)^2\right] - K \left\{ e^{i
  \frac{a}{f_a}} \det \left[\bar{\psi_j}\left(1 +
  \gamma^5\right)\psi_j\right] \right. \nonumber\\ &\left. +e^{- i \frac{a}{f_a}} \det
\left[\bar{\psi_j}\left(1 - \gamma^5\right)\psi_j\right] \right\} - G_{V} \left(\bar{\psi_j}
\gamma^{\mu} \psi_j\right)^2 ,
\label{lagr}
\end{align}
where the first, second and last terms are the usual NJL-type ones for the quarks, including scalar, pseudoscalar and vector interaction terms. $\psi_j$ are the Dirac fields for the quarks, $\gamma^\mu$ the Dirac matrices, $m_0^j$ the current quark masses, $\lambda^b$ the Gell-Mann matrices,
and $G_s$ and $G_V$ are, respectively, the coupling constants for the
scalar/pseudoscalar and the vector interactions. The third term
in Eq.~(\ref{lagr}) represents the axion contribution, i.e., the
interaction between the axion field $a$ and the quarks (with strength
$K$), through a chiral rotation by the angle $a/f_a$, $f_a$ being the
axion decay constant.  Within the mean-field approximation, we can
effectively replace the interactions with corresponding
condensates. 

Since we are interested in studying axion effects on the stability
of fully evolved hybrid NSs, which are equilibrated with respect to the weak force, several conditions can be imposed. These are effectively zero temperature ($T=0$), electric charge neutrality, meaning the number densities for the quarks
up, down,  and strange and for the electron should satisfy $\frac{2}{3}n_u - \frac{1}{3}\left(n_d + n_s\right) - n_e = 0$, along with the condition of $\beta$-equilibrium with a free Fermi gas of electrons ($\mu_e=-\mu_Q =-(\mu_u-\mu_d)$), and no constraint on strangeness ($\mu_S=0$), yielding $\mu_u = \frac{\mu_B}{3} + \frac{2}{3}\mu_Q$, and  $\mu_d = \mu_s = \frac{\mu_B}{3} - \frac{1}{3}\mu_Q$, where $\mu_u$, $\mu_d$, $\mu_s$, and $\mu_e$ are the chemical potentials for the quark flavors and electrons. The independent chemical potentials for the baryons $\mu_B$, charged $\mu_Q$, and strangeness $\mu_S$ correspond to the conserved quantities baryon number, electric charge, and strangeness (or lack of in our case).

{}For such a system, the thermodynamic potential reads
\begin{align}
\Omega &= \Omega_q + 2 G_s \sum_j \left(\sigma_{j}^{2} +
\eta_{j}^{2}\right) + 4 K \left(\sigma_u \sigma_d \sigma_s \cos
\frac{a}{f_a} \right. \nonumber \\ & \left. + \eta_u \eta_d \eta_s
\sin \frac{a}{f_a}\right) - 4 K \left[\cos \frac{a}{f_a} \left( \eta_u
  \eta_d \sigma_s + \eta_u \eta_s \sigma_d \right. \right. \nonumber
  \\ & \left. \left. + \eta_d \eta_s \sigma_u\right) + \sin
  \frac{a}{f_a}\left( \sigma_u \sigma_d \eta_s + \sigma_u \sigma_s
  \eta_d + \sigma_d \sigma_s \eta_u \right)\right] \nonumber \\ & -
G_V n^2 ,
\label{omega_axion}
\end{align}
where $\sigma_{j} = - \langle \bar{\psi_j} \psi_j \rangle$ and $\eta_{j}
= \langle \bar{\psi_j} i \gamma_5 \psi_j \rangle$ are the scalar and pseudoscalar quark condensates, respectively, and $n = \sum_j n_j=\sum_j
 \langle \psi_j^{\dagger} \psi_j\rangle$ is the total quark number density. 
The quark contribution $\Omega_q$ in
Eq.~(\ref{omega_axion}) is given by
\begin{align}
\Omega_q = - 2 N_c \!\!\sum_j \!\!\left[
  \int\limits_{\Lambda}\!\! \frac{d^3 k}{\left(2 \pi\right)^3} E_{k}^{j}
  +\!\! \int\limits_{k_{F}^{j}}\!\! \frac{d^3 k}{\left(2
    \pi\right)^3} \left(\tilde{\mu_j} - E_{k}^{j}\right)\right] ,
\end{align}
where $E_{k}^{j} = \sqrt{k^2 + \left.M^{j}\right.^2}$ with $M^{j} =
\sqrt{ \left.M_{s}^{j}\right.^2 + \left.M_{ps}^{j}\right.^2 }$
denoting the constituent quark masses. $M_{s}^{j}$ and $M_{ps}^{j}$
are the scalar and pseudoscalar contributions of the constituent mass,
given by the gap equations
\begin{align}
M_{s}^{j} &= m_{0}^{j} + 4 G_s \sigma_j + 2 K \left[\cos \frac{a}{f_a}
  \left( \sigma_k \sigma_l - \eta_k \eta_l \right) \right. \nonumber
  \\ & \left. - \sin \frac{a}{f_a} \left( \sigma_k \eta_l + \eta_k
  \sigma_l \right)\right],
\end{align}
\begin{align}
M_{ps}^{j} &= 4 G_s \eta_j - 2 K
\left[\cos \frac{a}{f_a} \left( \sigma_k \eta_l + \eta_k \sigma_l
  \right) \right. \nonumber \\ & \left. - \sin \frac{a}{f_a} \left(
  \eta_k \eta_l - \sigma_k \sigma_l \right)\right],
\end{align}
where $j,k,l = u,d,s$ (or cyclic permutations), $\tilde{\mu_j} = \mu_j - 2 G_V
n$ is the effective chemical potential, $\Lambda$ is the ultraviolet momentum cutoff and $k_F^j=
\sqrt{\tilde{\mu_j}^2-(M^j)^2}~\Theta(\tilde{\mu_j}^2-(M^j)^2)$ the Fermi momentum.

From the thermodynamic potential given by Eq.~(\ref{omega_axion}),
we can now find the physical values for the condensates  $\sigma_j$,
$\eta_j$, and $n$ by solving the appropriate gap equations $\frac{\partial \Omega}{\partial\sigma_j} = \frac{\partial
  \Omega}{\partial\eta_j}=  \frac{\partial \Omega}{\partial n} = 0$
%
%
, which also depend on the vacuum expectation value of the axion
background field $a$. Putting those physical values back in
Eq.~(\ref{omega_axion}), we obtain the effective thermodynamic
potential at finite quark chemical potential $\Omega (a,\mu)$. The
normalized thermodynamic potential is then defined by subtracting the vacuum value, $\Omega_N =
\Omega (a,\mu) - \Omega (a,0)$.  The total pressure, energy density,
and baryon number density are, respectively, given by
$p = - \Omega_N + \frac{\mu_{Q}^{4}}{12 \pi^2}$, $\epsilon = \Omega_N
+ \sum_j \mu_j n_j  + \frac{\mu_{Q}^{4}}{4 \pi^2}$ and $n_B
= \frac{1}{3} \sum_j n_j = \frac{1}{3 \pi^2}
\left(\left.k_{F}^{u}\right.^3 + \left.k_{F}^{d}\right.^3 +
\left.k_{F}^{s}\right.^3\right)$.
%

\begin{figure*}[!htb]
\centering
\begin{tabular}{c c c c}
\subfigure[]{\includegraphics[width=0.38\linewidth]{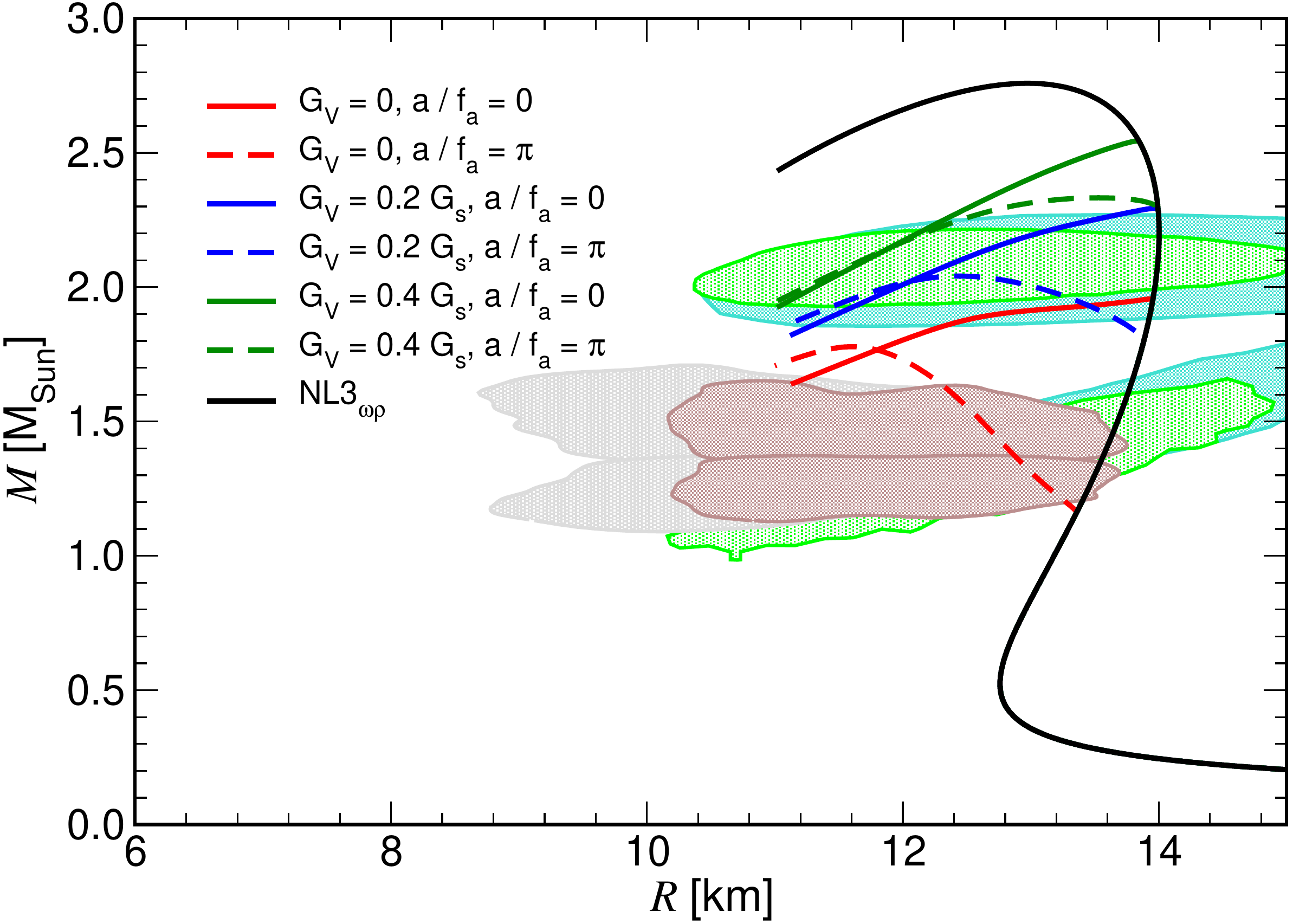}}
&
\subfigure[]{\includegraphics[width=0.38\linewidth]{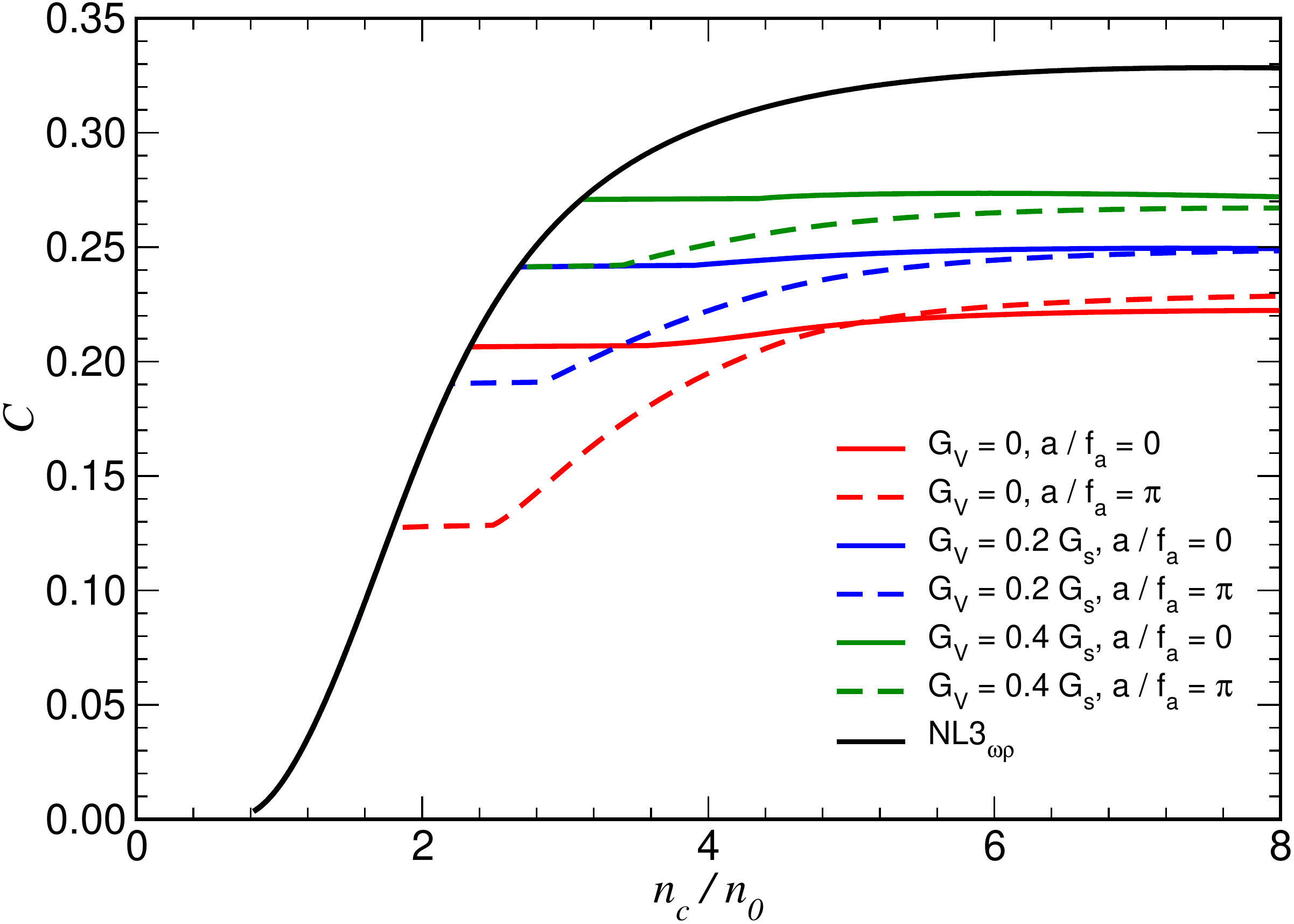}}
&
\\ \subfigure[]{\includegraphics[width=0.38\linewidth]{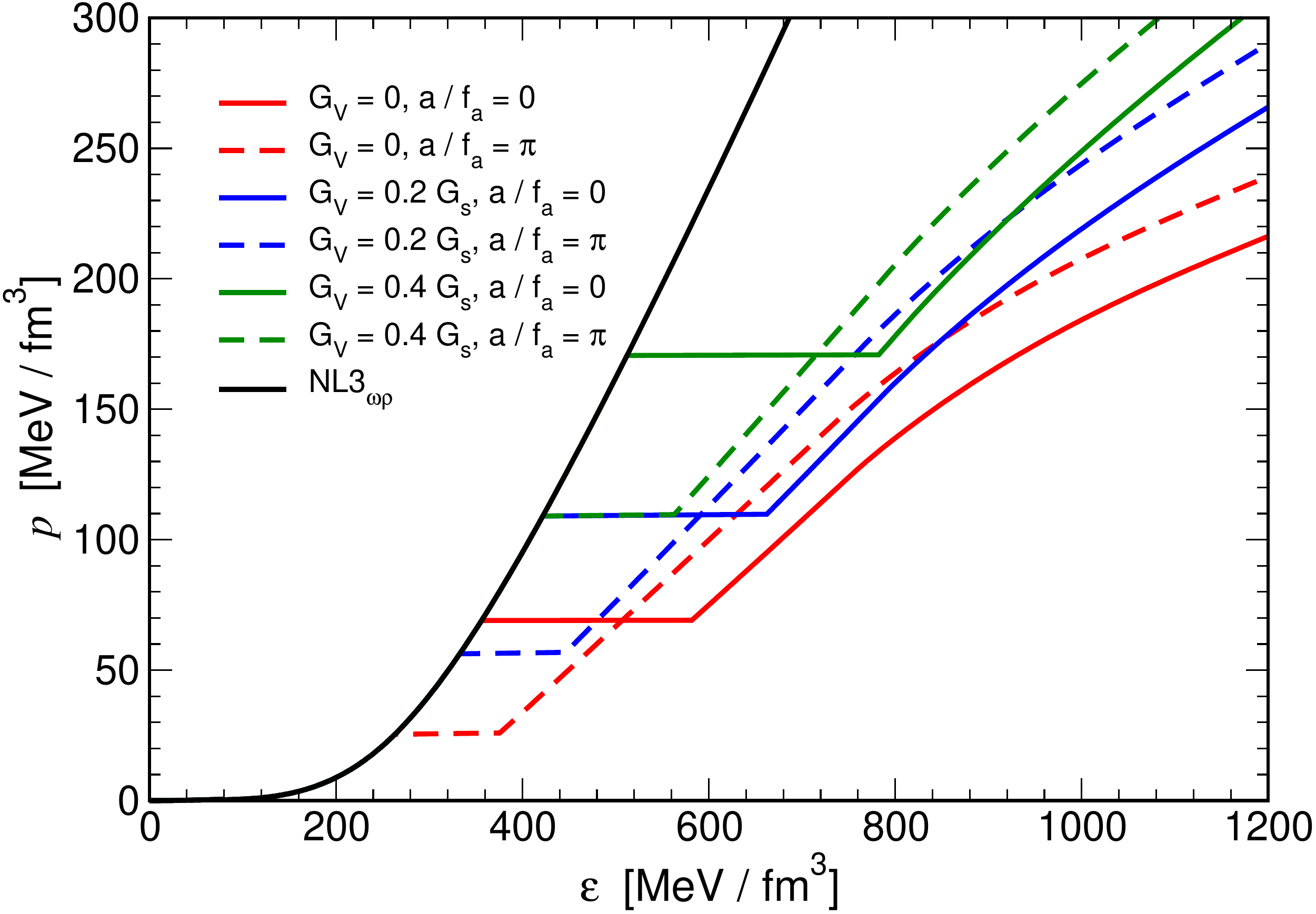}}
&
\subfigure[]{\includegraphics[width=0.38\linewidth]{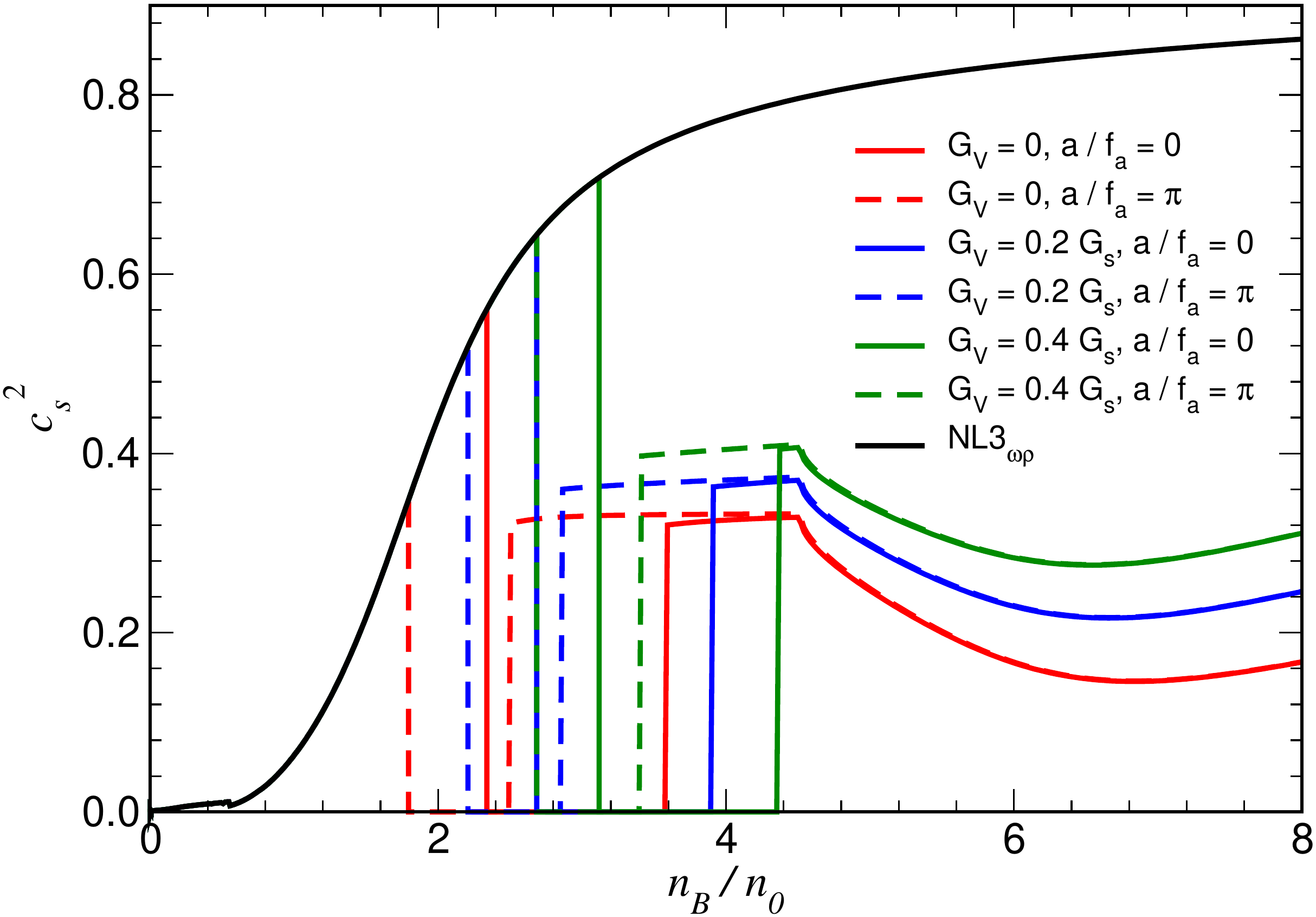}}
\end{tabular}
\caption{Results using the NL3$_{\omega \rho}$ crust: mass-radius relation
  (a), compactness vs. normalized central number density (b), equation of state (c), and speed of sound vs. normalized baryon number density (d)
  for different values of vector coupling $G_V$ and the scaled axion
  field $a/f_a$. The shaded regions are obtained
  from observational constraints from LIGO/Virgo (brown and gray) and NICER (two different shades of green) data~\cite{Tan:2021ahl}.}
\label{fig01:NL3wrcrust}
\end{figure*}

It is known, however, that finding an EoS describing stable pure quark
matter can be a challenging task, a problem that is aggravated with
the introduction of a repulsive vector interaction~\cite{Zacchi:2015oma}. This stems from the fact that stable pure quark matter must be more bound than iron at nuclear saturation density, which becomes more difficult for stiffer (larger $p(\varepsilon)$) EoS's. For this reason, one needs to consider a hadronic crust together with the
quark matter core. In this work, we will present results for the relativistic
NL3$_{\omega \rho}$ and CMF$_{\omega \rho,\omega^4}$ models. Both of them are in agreement with standard saturation properties and astrophysics observations (see regions shown in Fig.~1 for the latter). The NL3$_{\omega \rho}$~\cite{Horowitz:2001ya,Horowitz:2000xj,Lalazissis:1996rd} is a
nucleonic Walecka type model that contains the fewest ingredients that
allow hadronic matter to be in agreement with nuclear and
astrophysical observations, with $\omega \rho$ referring to a
mixed vector-isovector interaction that allows to reproduce smaller stars with lower tidal deformabilities, as measured by LIGO-Virgo \cite{LIGOScientific:2018cki}. The Chiral Mean Field CMF$_{\omega
  \rho,\omega^4}$ model~\cite{Dexheimer:2008ax,Clevinger:2022xzl} accounts for chiral symmetry
restoration, while also being in agreement with nuclear and
astrophysical observations. $\omega^4$ refers to a higher-order
vector interaction that allows to reproduce NSs with mass $M>2$ M$_{\rm{Sun}}$ including hyperon degrees of freedom. The complete EoSs also contain separate treatments
at very low density to account for the presence of nuclei. To describe nuclei, the CMF
model includes a unified EoS by Gulminelli and
Raduta~\cite{Gulminelli:2015csa} with effective Skyrme interaction of the type SkM proposed by L. Bennour
et. al.~\cite{Bennour:1989zz} and cluster energy functionals from
Danielewicz and Lee~\cite{Danielewicz:2008cm}. The NL3 includes the
Baym-Pethick-Sutherland (BPS) EoS~\cite{Baym:1971pw} and a
self-consistent Thomas-Fermi approach  with non-spherical pasta phases
\cite{Grill:2014aea}. The crust EoS's utilized in this work are available in the CompOSE
repository~\cite{Typel:2013rza,Oertel:2016bki,composewebsite}.

\begin{figure*}[!htb]
\centering
\begin{tabular}{c c c c}
\subfigure[]{\includegraphics[width=0.38\linewidth]{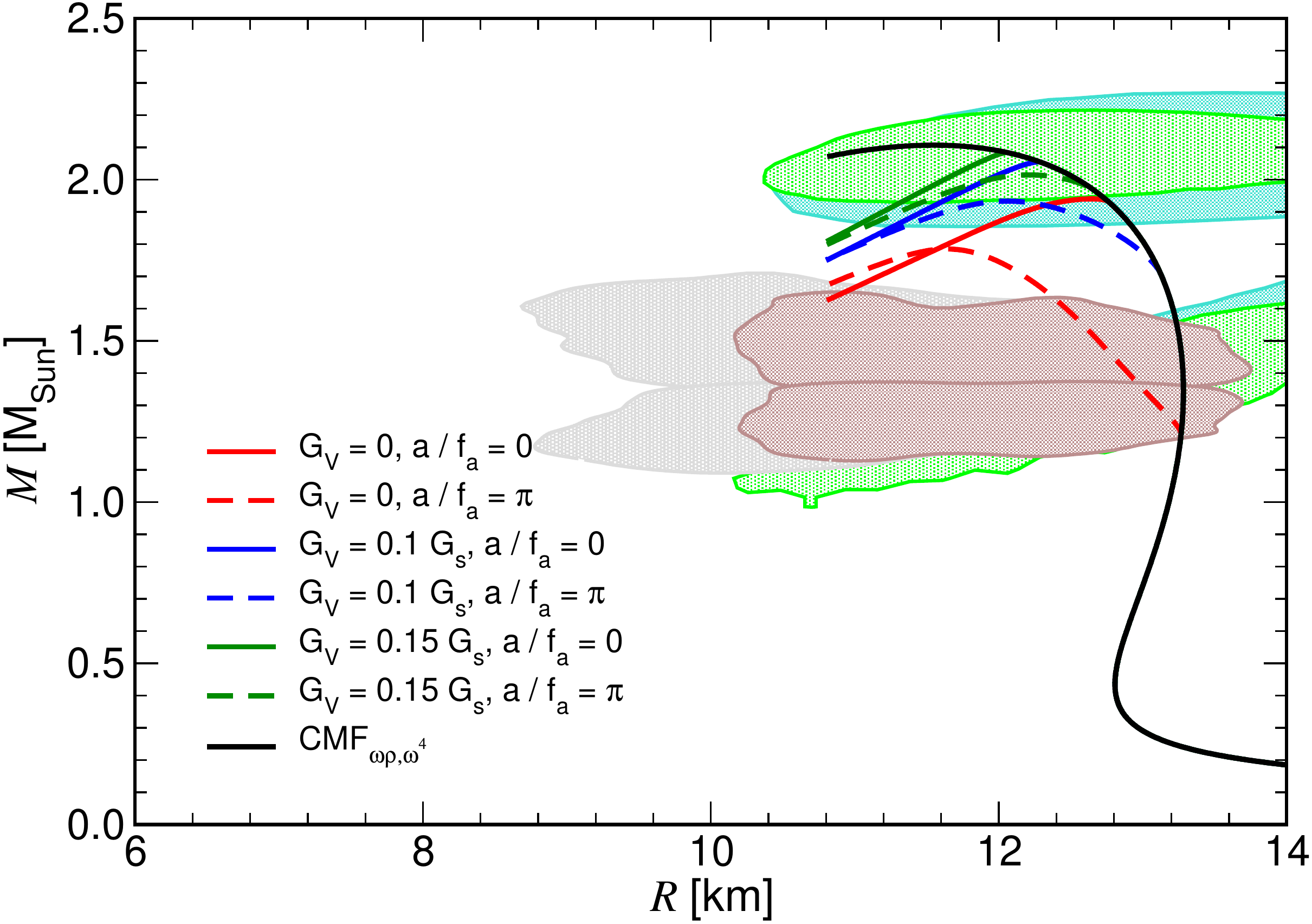}}
&
\subfigure[]{\includegraphics[width=0.38\linewidth]{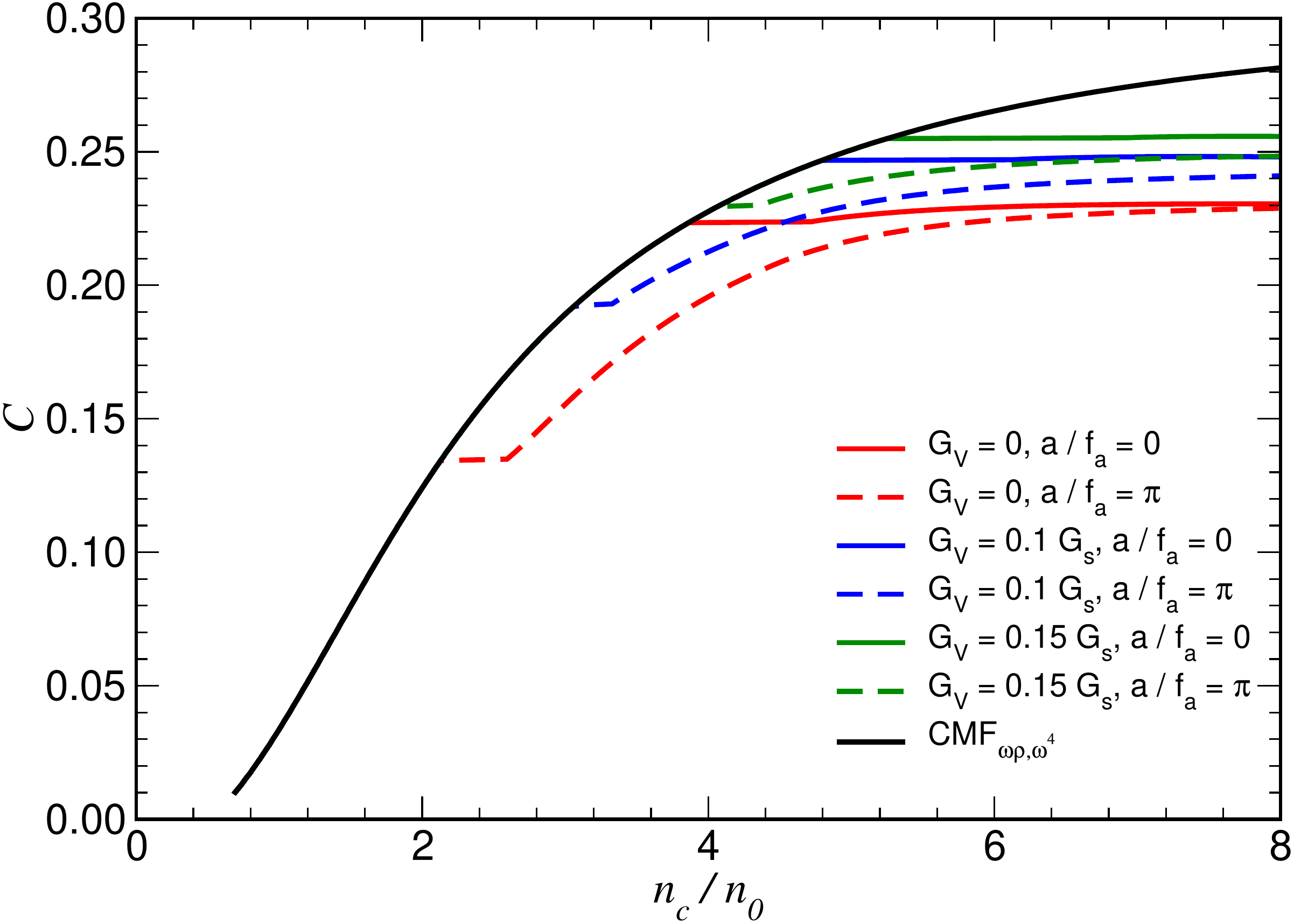}}
&
\\ \subfigure[]{\includegraphics[width=0.38\linewidth]{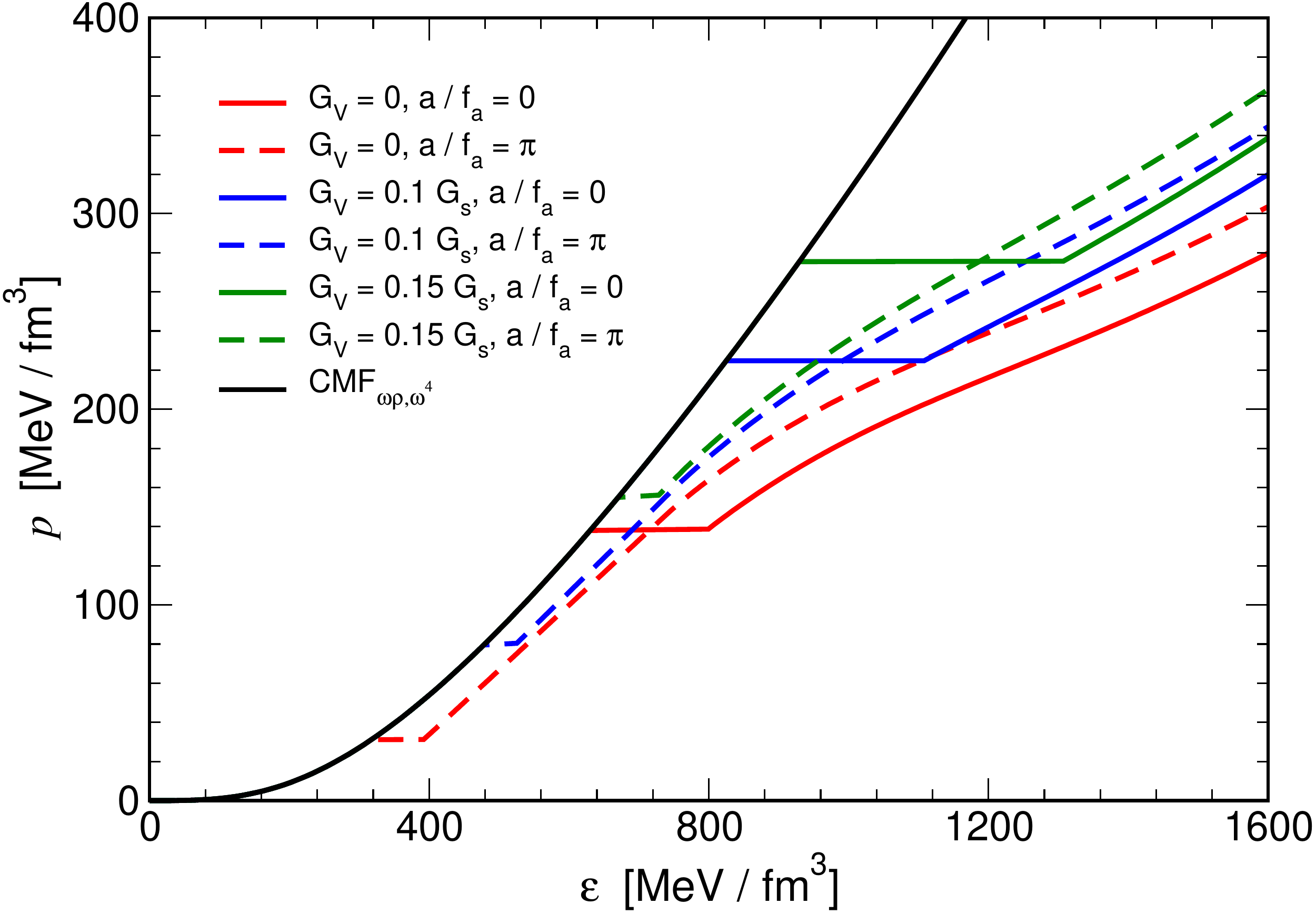}}
&
\subfigure[]{\includegraphics[width=0.38\linewidth]{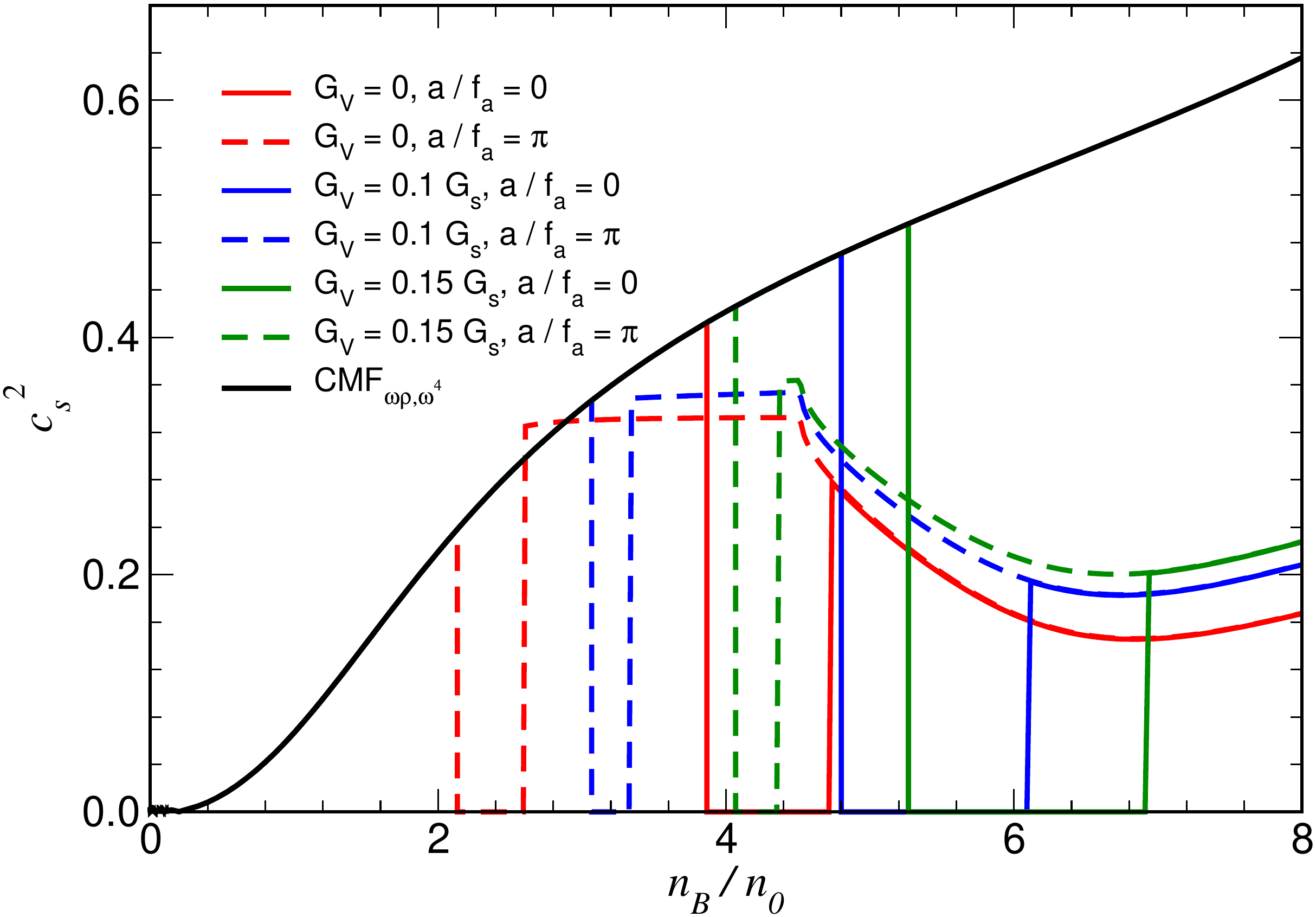}}
\end{tabular}
\caption{Same as Fig.~1 but using the CMF$_{\omega\rho,\omega^4}$ crust.}
\label{fig02:CMFwrcrust}
\end{figure*}

Finally, the mass-radius relation for a family of spherical, isotropic, static (or slowly rotating) stars is obtained solving
the Tolman-Oppenheimer-Volkoff (TOV)
equations~\cite{Oppenheimer:1939ne,Tolman:1939jz}.
In the numerical analysis discussed in the following, we consider the
parameters of our model to be $\Lambda = 631.4$ MeV, $G_s = 1.835 /
\Lambda^2$, $K = 9.29 / \Lambda^5$, $m^{u,d}_{0} = 5.5$ MeV, and
$m^{s}_{0} = 135.7$ MeV~\cite{Hatsuda:1994pi}. This set of parameters is traditionally used in the literature. The parameters are fixed by requiring that they satisfy experimentally measured properties of relevant quantities, e.g. the pion mass, the pion decay constant, the kaon mass, and the $\eta^\prime$ meson mass. They are representative enough to illustrate well-known features of the NJL model. The more flexible parameters $G_V$ and $a$ are varied widely in our analysis to study their effects on the various quantities we compute. In particular, we analyze the cases of $a/f_a = 0$ and $a/f_a = \pi$, such that the results can be shown in the absence of the axion effects and when these effects play a large role. Also, since $G_V$ has the same dimension as $G_s$ ($\sim 1/{\rm energy}^2$), it is natural to consider $G_V$ to be proportional to $G_s$, with the proportionality (dimensionless) factor taken here as a free parameter.


We start with the (complete) NL3$_{\omega \rho}$
model EoS for the hadronic crust. In Fig.~\ref{fig01:NL3wrcrust}, we
show (a) the obtained mass-radius relation for different stellar families
and corresponding behaviors for (b) the compactness $C = M/R$ (as a
function of central baryon number density normalized by the nuclear saturation
value $n_0=0.15$ $fm^{-3}$), (c) EoS $p(\varepsilon)$, and (d) speed of sound squared $c_s^2=dp/d\varepsilon$ (as a function of baryon number density normalized by the nuclear saturation value).
The kinks in panel (a), horizontal lines in panels (b) and (c), and $c_{s}^{2}=0$ in panel (d) (calculated as the derivative of panel (c)) are associated with a first-order phase transition between the hadronic crust and quark core. As a result of the first order phase transition,  there are jumps in first derivatives of the grand potential, such as number densities and energy density, which then manifest in the results shown in {}Fig.~\ref{fig01:NL3wrcrust}.
Stellar stability is guaranteed in the mass-radius diagram (starting from low density, bottom-right) until an extremum where the curve rotates counter-clockwise with increasing
central density. This can be derived from the Sturm-Liouville equation for radial stellar oscillations \cite{Alford:2017vca}.
{}From panel (a), accounting for the axion field
(through the non-zero ratio $a/f_a$) allows for stable branches of
hybrid NSs to exist, something not trivial when accounting for vector interactions $G_V\ne 0$. In our work, increasing
the value of the vector coupling allows the EoS to support stars of
higher masses. For value of the ratio $a/f_a = \pi$, we find stable stars with a maximum mass $M >
2$ M$_{\rm{Sun}}$ for $G_V = 0.2~G_s$ and $G_V = 0.4~G_s$.  Panel (b) of
Fig.~\ref{fig01:NL3wrcrust} shows the nature of the compactness with
varying values of $G_V$ and $a/f_a$ using $G=c=1$. For higher values of $G_V$, hybrid NSs are more compact, which agrees with our
observations from panel (a). The difference is that for $a/f_a\ne 0$ hybrid NSs are stable.

To better understand our findings described above, we discuss the
effect of $G_V$ and $a/f_a\ne 0$ on microscopic properties,
meaning the matter EoS. From panel (c) of Fig.~\ref{fig01:NL3wrcrust},
the transition from the hadronic crust to the quark matter core
happens at higher energy densities $\varepsilon$ and is stronger
(larger jump in $\varepsilon$ across the first-order phase transition)
for higher values of $G_V$. Increasing the ratio $a/f_a$ has the
effect of bringing the transition towards a smaller value of
$\varepsilon$ (for a specific $G_V$), in addition to making the jump
in $\varepsilon$ smaller at the transition. Both of these features are
known to help with stellar stability, as discussed in detail in Ref.~\cite{Alford:2013aca}. In panel
(d), we show the nature of the square of the speed of sound
$c_s^2$. The magnitude of $c_s^2$ can be understood as a measure of
stiffness of the EoS. While $G_V$ clearly turns the EoS
stiffer, accounting for the axion field does not modify the speed of
sound of quark matter away from the phase transition. Nevertheless, it
modifies significantly the phase transition region.  The bump in
$c_{s}^{2}$ in the quark matter phase, around $n = 3.6 \sim 4.4$ $n_0$,
happens when the $s$ quark starts to populate the system. Just before that, the speed of sound is close to being constant.

In Fig.~\ref{fig02:CMFwrcrust} we present our results obtained using a
different model EoS for the hadronic crust. In this case we choose the
(complete) CMF$_{\omega\rho,\omega^4}$ model and take the values of $G_V$ to be
$0$, $0.1$, and $0.15~G_s$. The four panels in Fig.~\ref{fig02:CMFwrcrust} indicate the same
quantities as Fig.~\ref{fig01:NL3wrcrust}. Looking at the black full
line of panel (d) for hadronic matter only (when compared to panel (d) of the previous figure),
it becomes clear that the CMF$_{\omega\rho,\omega^4}$ model EoS is very different
from the NL3$_{\omega\rho}$ model EoS. But, in spite of that, all our
conclusions from Fig.~\ref{fig01:NL3wrcrust} still hold. We still
reproduce hybrid and stable $2$ M$_{\rm{Sun}}$ stars (panel (a)), with the
difference that now such stars contain hyperons in the hadronic crust and a much smaller quark
core. Larger values of $G_V$ cannot be used in this case because they
would push the phase transition to densities not reached inside
NSs in the case of $a/f_a = \pi$. As a consequence, the compactness
of all the analyzed hybrid NSs are now more similar (panel (b)). The
energy density jumps across the phase transition are now narrower and
take place at larger energy densities (panels (c) and (d)).

In panel (a) of both Figs.~\ref{fig01:NL3wrcrust} and
\ref{fig02:CMFwrcrust}, it can be seen that we reproduce families
of stars that fulfill all astrophysical constraint shaded regions,
which were shown in Fig. 1 of Ref.~\cite{Tan:2021ahl} and
extracted from LIGO/Virgo gravitational wave observations
\cite{LIGOScientific:2018cki} and NICER X-ray
observations~\cite{Miller:2019cac,Miller:2021qha,Riley:2019yda,Riley:2021pdl}. The NICER regions, for both the observed low and the large mass stars, appear in pairs because they include results from two separate
collaborations that perform independent analyses. The two LIGO-Virgo regions correspond to two different approaches based on different prescriptions to access the EoS in a model-independent (to a degree) approach. Looking at the results
derived from $a/f_a\ne 0$ (dashed lines), which are the ones that
reproduce stable hybrid NSs, the lower mass regions are fulfilled by
either pure hadronic stars (black full lines) or hybrid NSs with
$G_V=0$. The issue is that the latter do not fulfill $M > 2$ 
M$_{\rm{Sun}}$,
as observed for the pulsar PSR
J0740+6620~\cite{Fonseca:2021wxt}. Because of that, we conclude that
within our framework, those are probably hadronic stars. This result
could be different had we used other hadronic crust models. But, more
interestingly, the higher mass regions are fulfilled by both hadronic
and hybrid NSs with $G_V \neq 0$, which is a consequence of the large radius range current observation constraints comprehend. 


In this paper, we have presented how a combination
of repulsive quark interactions and the presence of a non-vanishing
axion condensate, both implemented at the level of the NJL model, affect
the structure and stability of hybrid NSs. It has been known from recent
studies (see e.g.~\cite{Fattoyev:2017jql}) that in order to fulfill
the constraints on the tidal deformability for low mass NSs,
$M\sim 1.4$ M$_{\rm{Sun}}$, a soft EoS is necessary. At the same time,
to support NSs with $M \sim 2$ M$_{\rm{Sun}}$ against gravitational
collapse, a stiff EoS is required for intermediate to high densities. Our
results support such softening-stiffing of the EoS (followed by a phase transition to a stiff phase) within
a thermodynamical consistent approach, thanks to the combined effects
of vector interactions in the hadronic crust and the quark core, and the axion field condensate. The appearance of a ``bump'' in the speed of sound (as described above) can produce observables results that could be measured by LIGO/Virgo in the near future.

More specifically, the axion field modifies the quark EoS 
mainly around the deconfinement phase transition by weakening it and
bringing it to lower densities, thus allowing for a more extended
region for stability in the mass-radius diagram, as shown explicitly
in our results in Figs.~\ref{fig01:NL3wrcrust}(a) and \ref{fig02:CMFwrcrust}(a). The
axion field thus contributes non-trivially to allow for branches with
stable massive hybrid NSs, which cannot be achieved by the effects of the
vector interaction alone. In particular, our results show that for a magnitude of the axion field ratio $a/f_a = \pi$, stable stars with a maximum mass $M > 2$ M$_{\rm{Sun}}$
are allowed for $G_V = 0.2 - 0.4$ $G_s$. We expect that the results we
have presented in this paper to be complementary to the recent studies
concerning the effects of (bosonic) dark matter to the structure of compact
stars and help in further understanding those effects.


The authors thank Helena Pais for providing us with the $NL3$
equation of state. This work was partially
supported by CNPq, Grants  No. 309598/2020-6 and
No. 307286/2021-5; CAPES Finance  Code  001;  FAPERGS Grants Nos. 19/2551- 0000690-0 and
19/2551-0001948-3;  FAPERJ Grant
No. E-26/201.150/2021; NSF, Grants PHY1748621, MUSES
OAC-2103680, and NP3M PHY-2116686; PHAROS COST Action
CA16214; Guangdong Major Project of Basic
and Applied Basic Research No. 2020B0301030008; Alexander von Humboldt Foundation.


\end{document}